\documentclass[11pt]{article}
\usepackage{moriond,epsfig}

\def\be{\begin{equation}}
\def\ee{\end{equation}}
\def\bea{\begin{eqnarray}}
\def\eea{\end{eqnarray}}

\epsfverbosetrue

\def\3{\ss}

\newcommand{\tev}{{\rm Te}\kern-1.pt{\rm V}}
\newcommand{\gev}{{\rm Ge}\kern-1.pt{\rm V}}
\newcommand{\mev}{{\rm Me}\kern-1.pt{\rm V}}
\newcommand{\kev}{{\rm Ke}\kern-1.pt{\rm V}}
\newcommand{\gevsq}{\mbox{$\mathrm{{\rm Ge}\kern-1.pt{\rm V}}^2$}}
\newcommand{\gevmsq}{\mbox{$\mathrm{{\rm Ge}\kern-1.pt{\rm V}}^{-2}$}}

\newcommand{\mayor} {\mbox{\raisebox{-0.4ex}
{$\;\stackrel{>}{\scriptstyle \sim}\;$}}}

\newcommand{\sla}[1]{/\!\!\!#1}

\begin{document}
\vspace*{4cm}
\title{PROSPECTS FOR HIGGS SEARCHES VIA VBF AT THE LHC WITH ATLAS~\footnote{Talk given on Behalf of the ATLAS Collaboration at XXXIXth Rencontres de Moriond, La Thuile, Aosta Valley, Italy
March 28 - April 4, 2004.}}

\author{B.~Mellado}

\address{Physics Department \\
University of Wisconsin - Madison \\
   Madison, Wisconsin 53706 USA }

\maketitle\abstracts{ \noindent We report on the potential for the
discovery of a Standard Model Higgs boson with the vector boson
fusion mechanism in the mass range $115<M_H<500\,\gev/$c$^2$ with
the ATLAS experiment at the LHC. Feasibility studies at hadron
level followed by a fast detector simulation have been performed
for $H\rightarrow W^{(*)}W^{(*)}\rightarrow l^+l^-\sla{p_T}$,
$H\rightarrow\gamma\gamma$ and $H\rightarrow ZZ\rightarrow
l^+l^-q\overline{q}$. The preliminary results obtained here show a
large discovery potential in the range $115<M_H<300\,\gev/$c$^2$.
Results obtained with multivariate techniques are reported for a
number of channels. }

\section{Introduction}

In the Standard Model (SM) of electro-weak and strong
interactions, there are four types of  gauge vector bosons (gluon,
photon, W and Z) and twelve types of fermions (six quarks and six
leptons). These particles have been observed experimentally. The
SM also predicts the existence of one scalar boson, the Higgs
boson. The observation of the Higgs boson remains one of the major
cornerstones  of the SM. This is a primary focus of the ATLAS
Collaboration~\cite{LHCC99-14}.

The  Higgs at the LHC is produced predominantly via gluon-gluon
fusion~\cite{prl_40_11_692}. For Higgs masses, $M_H$,  such that
$M_H>100\,\gev/$c$^2$, the second dominant process is vector boson
fusion (VBF)~\cite{pl_136_196,pl_148_367}.

Early  analyses performed at the parton level indicated a strong
discovery potential with the decays $H\rightarrow W^{(*)}W^{(*)}$,
$\tau^+\tau^-, \gamma\gamma$ associated with to hard jets in the
range
$115<M_H<200\,\gev/$c$^2$~\cite{pr_160_113004,pl_503_113,pr_61_093005,JHEP_9712_005}.
The ATLAS collaboration has performed feasibility studies for
these decay modes including more detailed detector description and
the implementation of initial-state and final-state parton
showers, hadronization and multiple
interactions~\cite{SN-ATLAS-2003-024}.

Here, we present an update of the potential of observing the SM
Higgs boson via VBF with $H\rightarrow W^{(*)}W^{(*)}\rightarrow
l^+l^-\sla{p_T}$, where $\sla{p_T}$ stands for missing transverse
momentum carried by neutrinos. This analysis has been extended to
larger Higgs masses. Also, we investigated the prospects of
observing a SM Higgs boson with $H\rightarrow\gamma\gamma$ and
$H\rightarrow ZZ\rightarrow l^+l^-q\overline{q}$. Results obtained
with multivariate techniques are reported for a number of
channels. Finally, the status of the overall SM Higgs discovery
potential of the ATLAS detector is presented.

\section{Experimental Signatures}
\label{sec:expsig}

The VBF mechanism displays a number of distinct features, which
can be exploited experimentally to suppress SM backgrounds: Higgs
decay products are accompanied by two energetic forward jets
originating from incoming quarks and suppressed jet production in
the central region is expected due to the lack of color flow
between the initial state quarks. In this paper, tagging jets are
defined as the highest and next highest transverse momentum,
$P_T$, jets in the event. The tagging jets are required to be well
separated in pseudorapidity and to be in opposite hemispheres and
to have a large invariant mass.

The following decay chains have been considered in the analysis:
$H\rightarrow W^{(*)}W^{(*)}\rightarrow l^+l^-\sla{p_T}$, $H
\rightarrow \gamma \gamma$ and $H\rightarrow ZZ\rightarrow
l^+l^-q\overline{q}$. A number of relevant experimental aspects
have been addressed in detail
in~\cite{LHCC99-14,SN-ATLAS-2003-024} and will not be touched upon
here: triggering, lepton and photon identification, fake lepton
and photon rejection, jet tagging, central jet veto and b-jet veto
efficiencies. Details on specific event selections chosen for each
particular Higgs decay reported here are available
in~\cite{hep-ph_0401148}.

\section{\boldmath The $H\rightarrow W^{(*)}W^{(*)}\rightarrow l^+l^-\sla{p_T}$ Mode Associated with Two Hard Jets}
\label{sec:hww}

A study of this mode at hadron level followed by a fast simulation
of the ATLAS detector was first performed
in~\cite{ButtarHarperJakobs}. We report on a re-analysis over a
broader mass range $115<M_H<500\,\gev/$c$^2$. Additionally, the
treatment of the main background process is improved in the
present analysis and a mass dependent event selection has been
developed and implemented~\cite{hep-ph_0401148}. The proper
modelling of $t\overline{t}$ production associated with jets is
crucial to understanding the feasibility of this final state. For
this purpose, we used the MC package MC@NLO, which merges
consistently Next-to-Leading Order matrix elements with a parton
shower~\cite{JHEP_0206_029}.

\begin{table}[t]
\caption{Expected signal and background effective cross-sections
(fb) and the corresponding Poisson significance for the
$H\rightarrow W^{(*)}W^{(*)}\rightarrow l^+l^-\sla{p_T}$ decay
mode associated with two hard jets with 10\,fb$^{-1}$ of
integrated luminosity. A 10\% systematic uncertainty is applied to
all backgrounds when calculating the significance. }
\vspace{0.4cm}
\begin{center}
\begin{tabular}{c || c c c}
\hline \hline
$M_{H}(\gev/$c$^2)$ & $S$ & $B$  & Significance ($\sigma$) \\
\hline \hline
115     & 0.64   & 1.84    & 1.4\\
130     & 2.36   & 2.34    & 4.3\\
160     & 9.29   & 3.62    & 11.6\\
200     & 6.28   & 8.09    & 6.0\\
300     & 4.34   & 14.4    & 3.1\\
500     & 1.18   & 5.10    & 1.5\\
\hline \hline
\end{tabular}
\label{table:mcatnloSig}
\end{center}
\end{table}

The central jet veto survival probability for $t\overline{t}$
production is significantly lower than that reported
in~\cite{SN-ATLAS-2003-024}. However, this is compensated by a
lower rejection due to requiring two tagging jets. As a result,
the relative contribution to the background from $t\overline{t}$
production obtained here is similar to the one reported
in~\cite{SN-ATLAS-2003-024}. Table~\ref{table:mcatnloSig} reports
the expected signal and background effective cross-sections (in
fb) with the corresponding Poisson significance for 10\,fb$^{-1}$
of integrated luminosity. Simple event counting is used and a
$10\,\%$ systematic error on the background determination was
assumed. In order to implement the systematic errors we
incorporated~\cite{ATL-PHYS-2003-008,physics_03_12050} the
formalism developed in~\cite{nim_A320_331}. The $H\rightarrow
W^{(*)}W^{(*)}\rightarrow l^+l^-\sla{p_T}$ mode has a strong
potential in a wide rage of Higgs masses. A significance of or
greater than $5\,\sigma$ can be achieved with 30\,fb$^{-1}$ of
integrated luminosity for
$125<M_H<300\,\gev/$c$^2$~\cite{hep-ph_0401148}.

\section{\boldmath The $H\rightarrow\gamma\gamma$ Mode Associated with Two Hard Jets}
\label{sec:hgg}

This analysis was performed at parton level first
in~\cite{JHEP_9712_005}. The relevant background processes for
this mode are subdivided into two major groups. Firstly, the
production of two $\gamma$'s associated with two jets (real photon
production). Secondly, a sizeable contribution is expected from
events in which at least one jet is misidentified as a photon
(fake photon production). Despite the impressive jet rejection
expected to be achieved by the ATLAS detector~\cite{LHCC99-14}
($\mayor 10^3$ for each jet), the contribution from fake photons
will not be negligible due to the large cross-sections of QCD
processes at the LHC.

The contribution from the fake photon background has been severely
reduced due to the inclusion of the photon angular variables. The
contribution from this background, however, is important. The
normalization of the fake photon background is subject to sizeable
systematic uncertainties. This is partly due to the uncertainty on
the determination of the fake photon rejection
rate~\cite{LHCC99-14}. The signal significance expected with this
VBF mode alone reaches 2.2$\,\sigma$ for 30 fb$^{-1}$ of
integrated luminosity~\cite{ATL-PHYS-2003-036,hep-ph_0401148}. The
background estimation can be improved with the implementation of a
more realistic MC for the simulation of the real photon
background. This mode is considerably more sensitive to the
understanding of fake photon rejection than the inclusive
analysis~\cite{LHCC99-14}. A signal-to-background ratio of 0.5 or
better can be achieved, which gives room to less stringent
requirements on systematic errors. In real data the background
normalization can be obtained using side bands.

\section{\boldmath The $H\rightarrow ZZ\rightarrow l^+l^-q\overline{q}$ Mode Associated with Two Hard Jets}
\label{sec:hzz}

The main background corresponds to the production the QCD $Z+4j,
Z\rightarrow l^+l^-, l=e,\mu$. Diagrams with one or two EW boson
in the internal lines were neglected. The contribution from
$t\overline{t}$ is small and it is also neglected.

A feature specific to the mode under study is the additional
ambiguity in the definition of tagging jets introduced by the
presence of relatively hard jets produced from the decay of the
$Z$'s.  A search for two jets with an invariant mass close to the
$Z$ mass, $M_Z$, is performed. After reconstructing the $Z$
decaying hadronicaly, the event looks like a ``typical'' VBF
candidate.

Due to the application of kinematic fits, an average relative
invariant mass resolution of 2.5$\%$ can be obtained. A signal
significance of $3.75\,\sigma$ can be achieved for
$200<M_H<300\,\gev/$c$^2$ with 30\,fb$^{-1}$ of integrated
luminosity~\cite{hep-ph_0401148}. It should be noted that the
cross-sections for the main background reported here are subject
to large theoretical uncertainty. Fortunately, the background can
be determined from side bands for Higgs searches with
$M_H>200\,\gev/$c$^2$. The main source of systematic errors are
due to the energy scale uncertainty of hadronic jets. The energy
scale of hadronic jets should be known to $2\%$ in order to
achieve a $10\%$ error on the background normalization.

\section{Multivariate Analysis}
\label{sec:NN}

Results reported in~\cite{SN-ATLAS-2003-024} and the present paper
were based on classical cut analyses. Multivariate techniques were
used extensively in physics analyses, for instance, in LEP
experiments. Neural Networks (NN) are the most commonly used tools
in multivariate analyses. NN training has been performed on the
$H\rightarrow W^{(*)}W^{(*)}\rightarrow
l^+l^-\sla{p_T}$~\cite{ATL-PHYS-2003-007} and $H\rightarrow
\tau^+\tau^-\rightarrow l^+l^-\sla{p_T}$~\cite{NNVBFtautaull}
modes. NN training was performed with a relatively small number of
variables. It was required that these variables are infra-red safe
and their correlations do not depend strongly on detector effects.
The signal significance was calculated with a likelihood ratio
technique using the NN output as a discriminant
variable~\cite{ATL-PHYS-2003-008,physics_03_12050}. An enhancement
of approximately $30-50\,\%$ of the signal significance with
respect to the classical cut analysis was obtained for both modes
under consideration.

\section{Conclusions}
\label{sec:conclusions}

The discovery potential for the SM Higgs boson associated with two
hard jets in the range $115<M_H<500\,\gev/$c$^2$ has been
reported. An updated study at hadron level followed by a fast
detector simulation of the $H\rightarrow W^{(*)}W^{(*)}\rightarrow
l^+l^-\sla{p_T}$ mode has been presented: the main background,
$t\overline{t}$ associated with jets, has been modelled with the
MC@NLO program and the Higgs mass range has been extended to
$500\,\gev/$c$^2$. This mode has a strong potential: a signal
significance of more than $5\,\sigma$ can be achieved with
30\,fb$^{-1}$ of integrated luminosity for
$125<M_H<300\,\gev/$c$^2$. The discovery potential of the
$H\rightarrow\gamma\gamma$ and $H\rightarrow ZZ\rightarrow
l^+l^-q\overline{q}$ modes have also been reported. The discovery
potential of the $H\rightarrow\tau\tau$ modes have not been
updated here.

\begin{figure}[t]
\begin{center}
\epsfig{file=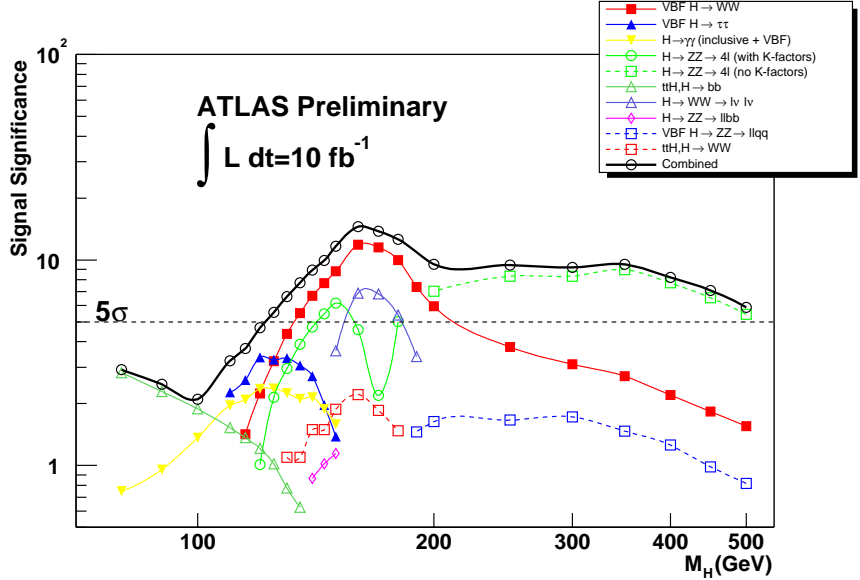, width=11.6cm}
\vspace{-0.2cm} \caption{Expected significance for ATLAS
(preliminary) as a function of Higgs mass for 10\,fb$^{-1}$ of
integrated luminosity.} \label{fig:discoveryPlot}
\end{center}
\end{figure}

The discovery potential of the modes presented here was combined
with results reported in past studies performed for the ATLAS
detector.  For the purpose of the combination the results reported
in~\cite{SN-ATLAS-2003-024} were used. Results from recent
studies~\cite{ATL-PHYS-2003-001,ATL-PHYS-2003-024,ATL-PHYS-2003-025},
which were not used in~\cite{SN-ATLAS-2003-024}, were added here.
Likelihood ratio techniques have been used to perform the
combination~\cite{ATL-PHYS-2003-008,physics_03_12050}. In order to
incorporate systematic errors, the formalism developed
in~\cite{nim_A320_331} was implemented. A 10\,\% systematic error
on the background estimation has been assumed for modes related to
VBF~\cite{SN-ATLAS-2003-024}. This issue needs to be addressed in
more detail using control samples with full simulation.
Figure~\ref{fig:discoveryPlot} displays the overall discovery
potential of the ATLAS detector with 10\,fb$^{-1}$ of integrated
luminosity. Results from NN based analyses and discriminating
variables have not been included in the combination. The present
study confirms the results reported
in~\cite{pr_160_113004,pl_503_113,pr_61_093005,JHEP_9712_005,SN-ATLAS-2003-024},
that the VBF mechanism yields a strong discovery potential at the
LHC in a wide range of the Higgs boson mass.

\section*{References}

\begin{thebibliography}{10}

\bibitem{LHCC99-14}
ATLAS Collaboration.
\newblock Detector and physics performance technical design report.
\newblock CERN-LHCC/99-14 (1999).

\bibitem{prl_40_11_692}
H.M.~Georgi, M.E.~Machacek, S.L.~Glashow and D.V. Nanopoulos.
\newblock 40:11, 1978.

\bibitem{pl_136_196}
R.~Cahn and S.~Dawson.
\newblock B136:196, 1984.

\bibitem{pl_148_367}
G.~Kane, W.~Repko  and W.~Rolnick.
\newblock B148:367, 1984.

\bibitem{pr_160_113004}
D.L. Rainwater and D.~Zeppenfeld.
\newblock D60:113004, 1999.

\bibitem{pl_503_113}
 N.~Kauer, D.L.~Rainwater, T.~Plehn and D.~Zeppenfeld.
\newblock B503:113, 2001.

\bibitem{pr_61_093005}
T.~Plehn, D.L.~Rainwater and D.~Zeppenfeld.
\newblock D61:093005, 2000.

\bibitem{JHEP_9712_005}
D.L. Rainwater and D.~Zeppenfeld.
\newblock JHEP {\bf 9712} (1997) 005.

\bibitem{SN-ATLAS-2003-024}
S.~Asai {\it et al.}
\newblock ATLAS Scientific Note SN-ATLAS-2003-024 (2003), submitted to EPJ,
  hep-ph/0402254.

\bibitem{hep-ph_0401148}
K.~Cranmer, Y.Q.~Fang, B.~Mellado, W.~Quayle, S.~Paganis and
Sau~Lan~Wu.
\newblock ATLAS Note ATL-PHYS-2004-005 (2003), hep-ph/0401148.

\bibitem{ButtarHarperJakobs}
C.~Buttar, K.~Jakobs and R.~Harper.
\newblock ATLAS Note ATL-PHYS-2002-033 (2002).

\bibitem{ATL-PHYS-2003-008}
K.~Cranmer, B.~Mellado, W.~Quayle and Sau~Lan~Wu.
\newblock ATLAS Note ATL-PHYS-2003-008 (2003).

\bibitem{physics_03_12050}
K.~Cranmer, W.~Quayle, B.~Mellado and Sau~Lan~Wu.
\newblock physics/0312050 (2003).

\bibitem{nim_A320_331}
R.D.~Cousins and V.L.~Highland.
\newblock A320:331, 1992.

\bibitem{JHEP_0206_029}
S.~Frixione and B.R.~Webber
\newblock JHEP {\bf 0206} (2002) 029

\bibitem{ATL-PHYS-2003-036}
K.~Cranmer, B.~Mellado, W.~Quayle and Sau~Lan~Wu.
\newblock ATLAS Note ATL-PHYS-2003-036 (2003),
\newblock hep-ph/0401088.

\bibitem{ATL-PHYS-2003-007}
K.~Cranmer, B.~Mellado, W.~Quayle and Sau~Lan~Wu.
\newblock ATLAS Note ATL-PHYS-2003-007 (2003).

\bibitem{NNVBFtautaull}
K.~Cranmer, B.~Mellado, W.~Quayle and Sau~Lan~Wu.
\newblock ATLAS Note, in preparation.

\bibitem{ATL-PHYS-2003-001}
E.~Gross, G.~Martinez, G.~Mikenberg  and L.~Zivkovic.
\newblock ATLAS Note ATL-PHYS-2003-001 (2003).

\bibitem{ATL-PHYS-2003-024}
J.~Cammin and M.~Schumacher.
\newblock ATLAS Note ATL-PHYS-2003-024 (2003).

\bibitem{ATL-PHYS-2003-025}
K.~Cranmer, B.~Mellado, W.~Quayle and Sau~Lan~Wu.
\newblock ATLAS Note ATL-PHYS-2003-025 (2003),
\newblock hep-ph/0307242.


\end{thebibliography}

\end{document}